# Excitonically-Induced Mechanisms of Inelastic Processes in Rare-Gas Solids

A.N. Ogurtsov[a,b,*] and G. Zimmerer[c]

[a]*Institute of Low Temperature Physics & Engineering of NASU, Lenin Avenue 47, Kharkov 61103, Ukraine*
[b]*National Technical University "KhPI", Frunse Street 21, Kharkov 61002, Ukraine*
[c]*Institut für Experimentalphysik, University of Hamburg, Luruper Chaussee 149, Hamburg 22761, Germany*

**Abstract**

The models of permanent lattice defect formation in rare-gas solids are discussed with a focus on a point defect formation in solid Ar. The processes of large-scale atomic displacements induced by electronic excitations were studied using the selective vacuum ultraviolet spectroscopy method. The coexistence of intrinsic excitonic mechanism and extrinsic Rydberg state induced excited-state mechanism of Frenkel defect formation was found.

*PACS:* 61.82.Ms, 71.35.-y, 78.55.Hx

*Keywords:* Rare-gas solids; Electronic excitations; Self-trapping; Frenkel pairs; Rydberg states

## 1. Introduction

The fundamental excitation of non-metallic solids by photons and beams of particles with kinetic energy below the threshold of knock-on of atoms from lattice sites (subthreshold excitation) is a powerful tool for materials modification [1]. The scission of the bonds, which stabilize the ground-state configuration by transferring the electronic excitation energy to the lattice, requires trapping or self-trapping of electronic excitations [2]. However the range of materials, which exhibit inelastic processes induced by electronic excitation, is limited to specific classes of materials, such as alkali halides, alkali earth fluorides and fused quartz [1].

Rare-gas solids (RGS), or atomic cryocrystals, are the model systems in physics and chemistry of solids, and a lot of information about electronic excitations in RGS has been documented in several books and reviews [2–6]. As a consequence of the closed electronic shells, solid Xe, Kr, Ar, and Ne are the simplest known solids of the smallest binding energy, $\varepsilon_b$, between atoms in the lattice. On the other hand, solid Ar and Ne have band-gap energies, $E_g$, exceeding that of LiF

---

[*] Corresponding author. Fax: 380-572-322370   *E-mail address*: ogurtsov@ilt.kharkov.ua (A.N. Ogurtsov)

and may be cited as the widest band-gap insulators. The excitonic mechanisms of subthreshold inelastic radiation-induced processes in solid Xe, Kr and Ne were studied recently [6–10]. In the present paper we include to overall picture of excitonically induced processes of large-scale atomic displacements in RGS the data on point defect formation in solid Ar following primary selective excitation by synchrotron radiation.

## 2. Experimental

The experiments were carried out at the SUPERLUMI-station at HASYLAB, DESY, Hamburg. The selective photon excitation was performed by near-normal incidence 2m primary monochromator with photon flux about $10^{12}$ photons/s at spectral resolution $\Delta\lambda=0.2$ nm. The VUV-luminescence analysis was performed both with low-resolution $\Delta\lambda=2$ nm Pouey high-flux monochromator equipped with a multisphere plate detector and with high-resolution $\Delta\lambda=0.1$ nm secondary 1 m near-normal incidence monochromator equipped with a position-sensitive detector. To considerably increase the incident photon flux we also irradiated the samples by zero-order of primary monochromator. The experimental setup and methods of sample preparation from vapor phase were described in detail elsewhere [10,11].

## 3. Results and discussion

The electronic properties of RGS have been under investigation since seventies and now the overall picture of creation and trapping of electronic excitations is basically complete. Because of strong interaction with phonons the excitons and holes in RGS are self-trapped, and a wide range of electronic excitations are created in samples: free excitons, atomic-like (A-STE) and molecular-like self-trapped excitons (M-STE), molecular-like self-trapped holes and electrons trapped at lattice imperfections [2–6]. The coexistence of free and trapped excitations and, as a result, the presence of a wide range of luminescence bands with well studied internal structure in the emission spectra of RGS enable one to reveal the energy relaxation channels and to detect the elementary steps in lattice rearrangement.

In the row Xe, Kr, Ar, Ne the atomic plarizability decreases as well as the crystal electron affinity; the latter becomes negative in the case of solid Ar and Ne [2–6]. As a consequence, the main channel of exciton self-trapping in Xe and Kr is the formation of M-STE, which is equivalent to a molecular dimer $R_2^*$ imbedded into the host lattice (R=rare gas atom). In the opposite case of solid Ne the negative electron affinity results in strong repulsion of the Rydberg orbital of the lowest $2p^53s$ excitation with the closed shell of the nearest neighbour atoms, which are pushed outwards. Such local lattice rearrangement leads to formation of A-STE, which is equivalent to an excited atom R* into the cavity (or "bubble"). In the intermediate case of solid Ar both M-STE and A-STE coexist. One can expect that the modification of the lattice may be made either by M-STE or by excitation of Rydberg atomic states. Since the energy of electronic excitation is transferred into kinetic energy of atomic motion over a unit cell, formation of three-, two-, or one-dimensional defects is ruled out. Only the point radiation defects, *viz.*, Frenkel pairs may emerge in the bulk of the crystal.

The most prominent feature in excitonic luminescence of Xe, Kr and Ar – the so-called *M*-band – is formed by $^{1,3}\Sigma_u^+ \rightarrow ^1\Sigma_g^+$ transitions in $R_2^*$ excimer. Each of the *M*-bands can be well approximated by two Gaussians: low energy subband $M_1$ and high energy one $M_2$ [8]. The subband $M_2$ is dominant in the luminescence of more perfect samples. The spectra of samples with a great number of initial defects are mainly determined by the component $M_1$. This suggests that the subband $M_2$ is emitted by the excitons which are self-trapped in the regular lattice while the component $M_1$ is emitted by the centers which are populated during trapping that occurs with the lattice imperfections involved.

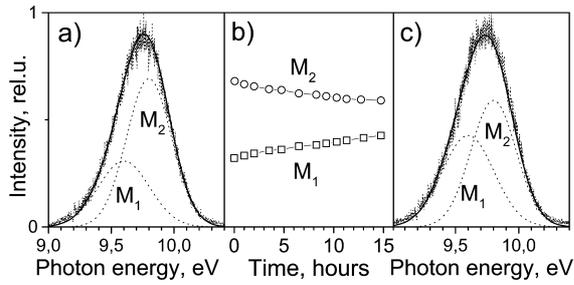

Fig. 1. Evolution of *M*-band of solid Ar under selective photoexcitation. (a) – *M*-band before irradiation, (b) – dose dependences of $M_1$ and $M_2$ subbands, (c) – *M*-band after irradiation.

Fig.1 shows the shape of the *M*-band of solid Ar before (a) and after (c) irradiation by photons with the energy $E_{exc}$=13.57 eV in the region of exciton band Γ(3/2), *n*=2 during 15 hours at *T*=7 K (for solid Ar $E_g$=14.15 eV). All spectra were normalized to exclude the variation of luminescence intensity because of sample sputtering by desorption and surface contamination by residual gases. Deconvolution into two subbands $M_1$ and $M_2$ shows that the intensity of $M_2$ subband, associated with self-trapping of excitons in regular lattice, diminishes while the intensity of $M_1$ subband, originating from molecular type centers in defect position, increases during irradiation time. Dose dependence for both components is plotted in Fig.1(b). From these data one can conclude that defect formation follows primary excitation of triplet Γ-excitons, and precursor states are the states of M-STE. The high quantum yield of molecular luminescence allows one to neglect non-radiative transitions, and the population of antibonding $^1\Sigma_g^+$ ground molecular state is usually considered as a main source of kinetic energy for a large-scale movement of atoms finishing in the Frenkel defects or desorption of atoms in the ground state – ground-state (GS) mechanism. On the other hand, the processes of formation of A-STE and M-STE centers themselves are accompanied by a considerable energy release to the crystal lattice, which also exceeds the binding energy $\varepsilon_b$ [6,11]. As in the case of solid Xe and Kr [6,7] the excited-state (ES) mechanism of M-STE to Frenkel pair conversion in solid Ar is supposed to occur by displacement of M-STE from centrosymmetric position in the <110> direction followed by reorientation to the <100> direction to stabilize the defect.

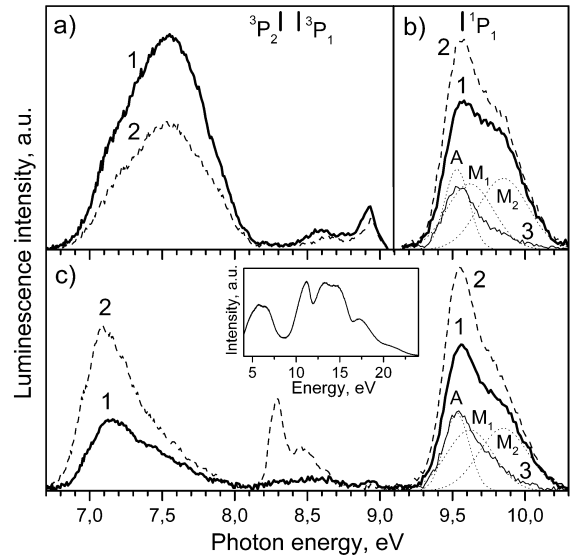

Fig. 2. Evolution of the luminescence spectrum of Xe doped solid Ar: (a) and (b) – as a result of selective photoexcitation with $h\nu$=9.2 eV; (c) – as a result of irradiation by zero-order of primary monochromator. Thick solid curves 1 – before irradiation; dashed curves 2 – after irradiation; thin solid curves 3 – difference of correspondent 2 and 1 curves. Inset: spectral distribution of photon flux from primary monochromator.

Recently the Frenkel-pair formation induced by excitation of Rydberg states of atomic-like centers was studied both by exciton self-trapping in solid Ne [9] (intrinsic mechanism) and by trapping of exciton at Ar impurity in Ne matrix [12] (extrinsic mechanism). In solid Ar the luminescence quantum yield from A-STE is approximately $10^4$ times lower than from M-STE and to test the possibility of defect formation by Rydberg electronic excitations we used the selective photoexcitation of the lowest dipole allowed extravalence Rydberg $^2P_{3/2}6s[3/2]_1$ state of Xe guests deep inside the Ar matrix. Figures 2(a) and 2(b) show the modification of luminescence spectrum of 1:1000 Xe:Ar sample as a result of $^1S_0 \to {}^3P_1$ selective excitation of Xe impurity atoms during 55 min by photons with $h\nu=9.2$ eV at $T=9$ K. The spectra of atomic and molecular emission of impurity Xe (Fig.2(a)) were measured at the beginning (curve 1) and before finishing (curve 2) of the irradiation. The spectra of $M$-band of Ar matrix (Fig.2(b)) were measured at excitation energy $E_{exc}=12.06$ eV (energy of $\Gamma(3/2)$ $n=1$ exciton) before (curve 1) and after (curve 2) irradiation. The intensity of the $M$-band increases at the expense of the "defect" $M_1$-subband (Fig.2(b), curve 3). Irradiation of the non-exposed part of the same sample during 15 min by the zero-order of primary monochromator, which excites mainly the host matrix (Fig.2(c) inset) and results in Frenkel defect formation *via* self-trapping of excitons, produces the similar change of $M$-band (Fig.2(c), curve 3). This confirms the extrinsic ES-mechanism of point defect formation by selective excitation of Rydberg states of impurity atoms. In solid Ar, as in the case of solid Ne, the strong repulsion of the Rydberg electron with a closed shell of surrounding atoms induces a substantial local lattice rearrangement, which leads to a "bubble" formation around the excited atom.

The decrease in intensity of impurity molecular emission (Fig.2(a)) and growth of quasi-$^1P_1$ luminescence band from Xe impurity atoms (Fig.2(b), dotted curve A) reflects the processes of "sputtering" of Xe clusters in the Ar matrix, movement of Xe atoms outward the Xe inclusions and trapping of impurities at radiation defects. The excitation of the sample basically *via* matrix excitations (Fig.2(c)) induces the diffusion and aggregation of impurities and trapping of Xe atoms at Frenkel defects as well.

The ES-mechanism of Frenkel-pair formation as a result of excitation of Rydberg atomic states is confirmed by recent molecular dynamics calculations [13,14], theoretical and experimental investigations of structural dynamics of electronically excited XeAr$_N$ clusters [15,16]. After the bubble formation the surrounding ground state atoms appear to have moved to the second shell. It was found that the second-nearest neighboring vacancy-interstitial pairs could create the permanent defects, which remain in the lattice after exciton annihilation [14].

In summary, we see that in solid Ar, as well as in all RGS, the selective excitation of excitons by photons of energies below $E_g$ results in accumulation of Frenkel-pairs, which is a direct proof of the excitonic nature of intrinsic ES-mechanism of defect formation *via* self-trapping of excitons. In addition in solid Ne and Ar an extrinsic ES-mechanism of defect formation *via* excitation of Rydberg states of atomic-like centers may considerably alter the total yield of inelastic radiation-induced processes. The support of DFG grant 436 UKR 113/55/0 is gratefully acknowledged.